\documentclass[aps,prb]{revtex4}
\usepackage{graphicx}
\usepackage{subfigure}  % use for side-by-side figures
\usepackage{amsmath} 
\usepackage{amssymb}
\begin{document}
%%%%%%%% Author macros begin:
\newcommand{\pderiv}[2]{\frac{\partial #1}{\partial #2}}
%%%%%%%% Author macros end
%
%
%
\title{Quantum and classical criticalities in the 
frustrated two-leg Heisenberg ladder}

\author{Mohamed Azzouz}
\email[Electronic Address: ]{mazzouz@laurentian.ca} 
\author{Brandon W. Ramakko}
 
\affiliation{Laurentian University, Department of Physics,
Ramsey Lake Road, Sudbury, Ontario P3E 2C6, Canada.}
  \date{August 21, 2007}

%\date{August 15, 2007}
\begin{abstract}
This talk was about 
the frustration-induced criticality 
in the antiferromagnetic Heisenberg model on the two-leg ladder 
with exchange interactions along the chains, rungs, and 
diagonals, and also about the effect of thermal fluctuations on 
this criticlity. 
The method used is the bond mean-field theory, which is
based on the Jordan-Wigner transformation 
in dimensions higher than one. 
In this paper, we will summarize the main
results presented in this talk, and report on new results 
about the couplings and temperature dependences of the 
spin susceptibility. %This talk was presented at Theory Canada 3 in 2007.
\end{abstract}
\maketitle

\newpage

\section{Introduction}

We use the bond-mean-field theory 
(BMFT), which is based 
on the Jordan-Wigner (JW) transformation,
to study the quantum 
criticality phenomenon in the frustrated 
antiferromagnetic (AF) two-leg Heisenberg ladder, 
and the effect of temperature on this criticality
\cite{Azz1,Azz2,Azz3,Azz5,ramakko2007}.
This method has been 
applied to the Heisenberg single chain, two-leg ladder, and three-leg 
ladder without frustration with excellent results \cite{azzouz2005}.
When the diagonal interaction is varied the two-leg ladder system 
can undergo a quantum phase transition between two of three distinct 
non-magnetic quantum spin liquid states; the N\'eel-type (N-type) state, 
ferromagnetic-type rung (R-type) state, and ferromagnetic-type chain 
(F-type) state. 
These states are characterized by ferromagnetic spin 
arrangements along the diagonals, rungs, or chains respectively. 
The BMFT is a mean-field theory that is based on the spin bond parameters. 
The latter are related to 
the spin-spin correlation function 
$\langle S_{i}^-S_{j}^+\rangle$, with 
$i$ and $j$ labeling two adjacent sites 
in the direction where this correlation 
function is calculated. All quantities 
$\langle S_i^\alpha\rangle$, 
with $\alpha=x,\ y,\ z$, are zero 
in BMFT, implying the absence of any 
sort of long-range magnetic order. 

\begin{figure}
\includegraphics[height=2.2cm]{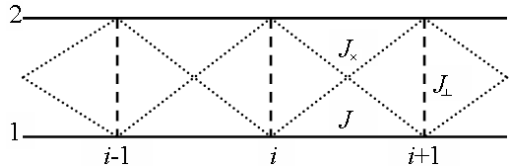}
\centering
\caption{The two-leg ladder showing the couplings 
along the chains, rungs, and diagonals is displayed.}
\label{fig:Two leg ladder}
\end{figure}

The Hamiltonian for the 
spin-$\frac{1}{2}$ two-leg ladder with diagonal 
interactions is written as
\begin{equation}
\label{eq:Ham}
H = J \sum_{i}^N \sum_{j=1}^2 \textbf{S}_{i,j}\cdot \textbf{S}_{i+1,j} +
J_\perp \sum_{i}^N 
\textbf{S}_{i,1}\cdot\textbf{S}_{i,2}+ J_{\times} 
\sum_{i}^N ( \textbf{S}_{i,1}\cdot\textbf{S}_{i+1,2} 
+ \textbf{S}_{i+1,1}\cdot\textbf{S}_{i,2}),
\end{equation}
where $J$ is the coupling along the chains, $J_\perp$ 
the coupling along the rungs, and $J_{\times}$ the coupling 
along the diagonals as seen in Fig.~\ref{fig:Two leg ladder}. 
The index $i$ labels the position of the spins along the 
two chains, each of which has $N$ sites, and $j$ labels the chains. 
As usual, $\textbf{S}_{i,j}$ is the spin operator.

The frustrated two-leg ladder has been studied numerically 
using the Ising and dimer expansions~\cite{Jd6}, the Lanczos 
diagonalization technique~\cite{Jd6,Jd11,Jd4,Jd8}, 
and the density-matrix renormalization group 
(DMRG)~\cite{Jd8,white1996,Jd5,Jd10,Jd3}. 
It has also been studied analytically 
using the bosonization~\cite{Jd6,Jd13}, 
the valence-bond spin wave theory~\cite{Jd7}, 
a non-perturbative effective-field 
theory~\cite{Jd9}, the non-linear sigma 
model~\cite{Jd12}, the reformulated 
weak-coupling field theory~\cite{starykh2004}, 
and the Lieb-Mattis theorem~\cite{hakobyan2007}. 
None of these analytical works 
addressed the issue of the phase diagram in all regimes including the 
weak, intermediate, and strong coupling limits. 
Within the BMFT, the phase diagram and its temperature
dependence can be easily examined in all these regimes.

This paper is organized as follows. 
In Sec.~\ref{sec:Method}, we explain how the BMFT, 
which is based on the 
JW transformation in dimensions higher than one, 
is applied to our Hamiltonian.  The
Quantum and classical critical behaviours, and the spin 
susceptibility are discussed in Sec. 
\ref{sec:Results}.  In Sec.~\ref{sec:Conclusion}, 
conclusions are reported.

\section{Method}

\label{sec:Method}

The JW transformation for the two-leg Heisenberg ladder 
is defined as~\cite{Azz3}
\begin{eqnarray}
\label{eq:2D JW}
S_{i,j}^{-} & = & c_{i,j}e^{i\phi_{i,j}}, \ \ \ 
S_{i,j}^{z}  =  n_{i,j}-1/2,\ \ \ \ 
n_{i,j}=  c_{i,j}^{\dag}c_{i,j}, \nonumber \\
\phi_{i,1} & = & \pi[\sum_{d=0}^{i-1}
\sum_{f=1}^{2}n_{d,f}]\ \ {\rm\ for\ chain\ 1,} \nonumber \\
\phi_{i,2} & = & \pi[\sum_{d=0}^{i-1}
\sum_{f=1}^{2}n_{d,f} + n_{i,1}] \ \ {\rm\ for\ chain\ 2}.
\end{eqnarray}
The $c_{i,j}^{\dag}$ operator creates a spinless 
fermion at site $(i,j)$, while $c_{i,j}$ annihilates 
one, and $n_{i,j}$ is the occupation number operator at that site. 
The phases $\phi_{i,j}$ are chosen so that at the 
the spin operators commutation relations are preserved.

After applying the JW transformation~(\ref{eq:2D JW}) 
to the Hamiltonian~(\ref{eq:Ham}) we get 
\begin{eqnarray}
\label{eq:Ham1}
H && =  \frac{J}{2} \sum_{i}^N (c_{i,1}^{\dag}e^{i\pi n_{i,2}}c_{i+1,1} 
+ c_{i,2}^{\dag}e^{i\pi n_{i+1,1}}c_{i+1,2} + \textrm{H.c.})  + 
\frac{J_\perp}{2} \sum_{i}^N (c_{i,1}^{\dag}c_{i,2} + \textrm{H.c.}) 
\nonumber \\
&&+ \frac{J_{\times}}{2} \sum_{i}^N 
(c_{i,1}^{\dag}e^{i\pi (n_{i,2}+n_{i+1,1})}c_{i+1,2} 
+ c_{i+1,1}^{\dag}c_{i,2} + \textrm{H.c.}) \nonumber
 + J \sum_{i}^N \sum_{j=1}^2 (n_{i,j} 
- \frac{1}{2})(n_{i+1,j} - \frac{1}{2}) \nonumber \\
&& + J_\perp \sum_{i}^N (n_{i,1} - \frac{1}{2})(n_{i,2} 
- \frac{1}{2}) + J_{\times}\sum_{i}^{N} [(n_{i,1} 
- \frac{1}{2})(n_{i+1,2} - \frac{1}{2}) \nonumber \\
&& + (n_{i+1,1} - \frac{1}{2})(n_{i,2} 
- \frac{1}{2})].
\end{eqnarray}
In BMFT, the interacting terms of the JW fermions are decoupled 
using the spin bond parameters. This approximation neglects 
fluctuations around the mean field points; 
$(O-\langle O\rangle)(O'-\langle O'\rangle) \approx 0$, 
where $O$ and $O'$ are any operators which are quadratic in 
$c^\dag$ and $c$~\cite{Azz2}. This yields
\begin{equation}
\label{eq:HF}
OO' \approx \langle O\rangle O' + O\langle O'\rangle 
- \langle O\rangle \langle O'\rangle.
\end{equation}
To apply BMFT we introduce three mean-field bond parameters; 
$Q$ in the longitudinal direction, $P$ in the transverse 
direction, and $P'$ along the diagonal. These can be 
interpreted as effective hopping energies for the JW 
fermions~\cite{Azz1} in the longitudinal, transverse 
and diagonal directions, respectively:
\begin{equation}
\label{eq:bondparameters}
Q  =  \langle c_{i,j}c_{i+1,j}^{\dag}\rangle,\qquad P  
=  \langle c_{i,j}c_{i,j+1}^{\dag}\rangle,\qquad P'  
=  \langle c_{i+1,j}c_{i,j+1}^{\dag}\rangle.
\end{equation} %
Keeping in mind that there is no long-range 
order~\cite{Mermin} so that $\langle S_{i,j}^{z}\rangle 
= \langle c_{i,1}^{\dag}c_{i,1}\rangle -1/2 = 0$, 
the Ising quartic terms in equation~(\ref{eq:Ham1}) 
can be decoupled using the Hartree-Fock 
approximation~(\ref{eq:HF}), and the bond 
parameters~(\ref{eq:bondparameters}) as follows:
\begin{eqnarray}
\label{eq:decouple Ising}
\left(c_{i,1}^{\dag}c_{i,1}\! 
- \!\frac{1}{2}\right)\left(c_{i+1,1}^{\dag}c_{i+1,1} 
\!-\! \frac{1}{2}\right) & \approx  & Qc_{i,1}^{\dag}c_{i+1,1} 
+ Q^*c_{i+1,1}^{\dag}c_{i,1} + |Q|^2,
\end{eqnarray}
\begin{figure}
\includegraphics[height=4.0cm]{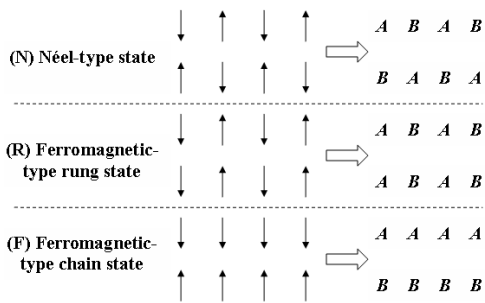}
\centering
\caption{In the left panel, the three possible 
ground states of the system in the Ising limit, 
namely the N\'eel state, the ferromagnetic chain 
state, and the ferromagnetic rung state are drawn. In the right
panel, the labeling of sublattices corresponding 
to the short-range spin orders that replace the long-range ones are
shown for the Heisenberg limit.}
\label{fig:Jd1}
\end{figure}
for the Ising interaction along the chain 1. Similar equations
can be obtained for the interactions along the other directions. 
Next, we write the Hamiltonian (\ref{eq:Ham1}) using the three different 
spin configurations in the right panel of Fig. \ref{fig:Jd1}. 
These configurations are instantaneous (not static) 
configurations in which adjacent spins in any direction 
keep on average the same relative orientations with 
respect to each other, but fluctuate globally on a 
time scale determined by the strongest coupling 
constant so that any kind of long-range magnetic 
order is absent. These fluctuations are a consequence 
of the spin quantum fluctuations. The three competing 
configurations of Fig. \ref{fig:Jd1} lead to three 
different quantum gapped spin liquid states, 
each characterized by its own short-range spin correlations and symmetry.
We also choose to place an alternating phase of $\pi$ along the chains 
so that the phase per plaquette is $\pi$ \cite{Affleck}. 
This phase configuration is used 
to get rid of the phase terms in the Hamiltonian. We also set $Q_{i,j} 
= Qe^{i\Phi_{i,j}}$ where $Q$ is site independent \cite{Azz5}. Here 
$\Phi_{i,j}$ is the phase of the bond along the chain; i.e.,
if $\phi_{i,j} = 0$ on a given bond, then it is equal to $\pi$
on the adjacent ones. This is necessary in order to recover the proper 
result in the limit $J_{\times}$ and $J_{\perp}$ becoming zero, 
in which we get an energy spectrum comparable to that of des Cloiseaux 
and Pearson~\cite{Cloiseaux} for the spin excitation
in a single Heisenberg chain, $E(k) = \frac{\pi}{2}J\left|\sin k\right|.$

For each state the Hamiltonian is written in the 
Nambu formalism and the matrix is diagonalized 
to obtain four eigenenergies (for each of the states). 
The details can be 
found in Ref.~\cite{ramakko2007}. The eigenenergies 
for the N, F, and R-type states are given respectively by
\begin{eqnarray}
E_{N}(k) \!&\! =\! &\! \pm \!J_{\times1}\cos k \pm 
\sqrt{J_{1}^{2}\sin^{2} k + \frac{J_{\perp 1}^{2}}{4}}, \nonumber \\
%\qquad
E_{F}(k)  \!&\! =\! &\!  \pm J_{1}\cos k \pm \sqrt{J_{\times1}^{2}\sin^{2}k 
+ \frac{J_{\perp 1}^{2}}{4}}, \nonumber \\
E_{R}(k) \!&\! =\! &\!  \pm \frac{J_{\perp 1}}{2}\pm 
\sqrt{J_{1}^{2}\sin^{2}k 
+ J_{\times1}^{2}\cos^{2}k},
\end{eqnarray}
where $
J_{1} =  J(1+2Q)$, $ J_{\perp1} =  J_{\perp}(1+2P)$, 
and $J_{\times1} =  J_{\times}(1+2P')$.
The free energy per site is
\begin{equation}
F  =  JQ^2 + \frac{J_{\perp}P^{2}}{2} + J_{\times}P'^{2} 
- \frac{k_{B}T}{4N} \sum_{k}\sum_{p=1}^{4}\ln[1 + e^{-\beta E_{p}(k)}],
\end{equation}
where $E_p(k)$ is one of the four eigenenergies in any state. 
The minimization of the free energy with respect to the bond parameters 
leads to the following set of self-consistent equations:
\begin{eqnarray}
\label{eq:sc}
Q & = & -\frac{1}{8NJ}\sum_{k}\sum_{p=1}^{4}
\pderiv{E_{p}(k)}{Q}n_{F}[E_{p}(k)],  \nonumber \\
P  &=&  -\frac{1}{4NJ_{\perp}}\sum_{k}\sum_{p=1}^{4} 
\pderiv{E_{p}(k)}{P}n_{F}[E_{p}(k)],  \nonumber \\
P' & = & -\frac{1}{8NJ_{\times}}\sum_{k}\sum_{p=1}^{4} 
\pderiv{E_{p}(k)}{P'}n_{F}[E_{p}(k)], 
\end{eqnarray}
which are solved numerically, except in the high-temperature regime, where 
analytical results are obtained.

\section{Results}

\label{sec:Results}

\begin{figure}(a)
\subfigure{
\includegraphics[scale=0.25]{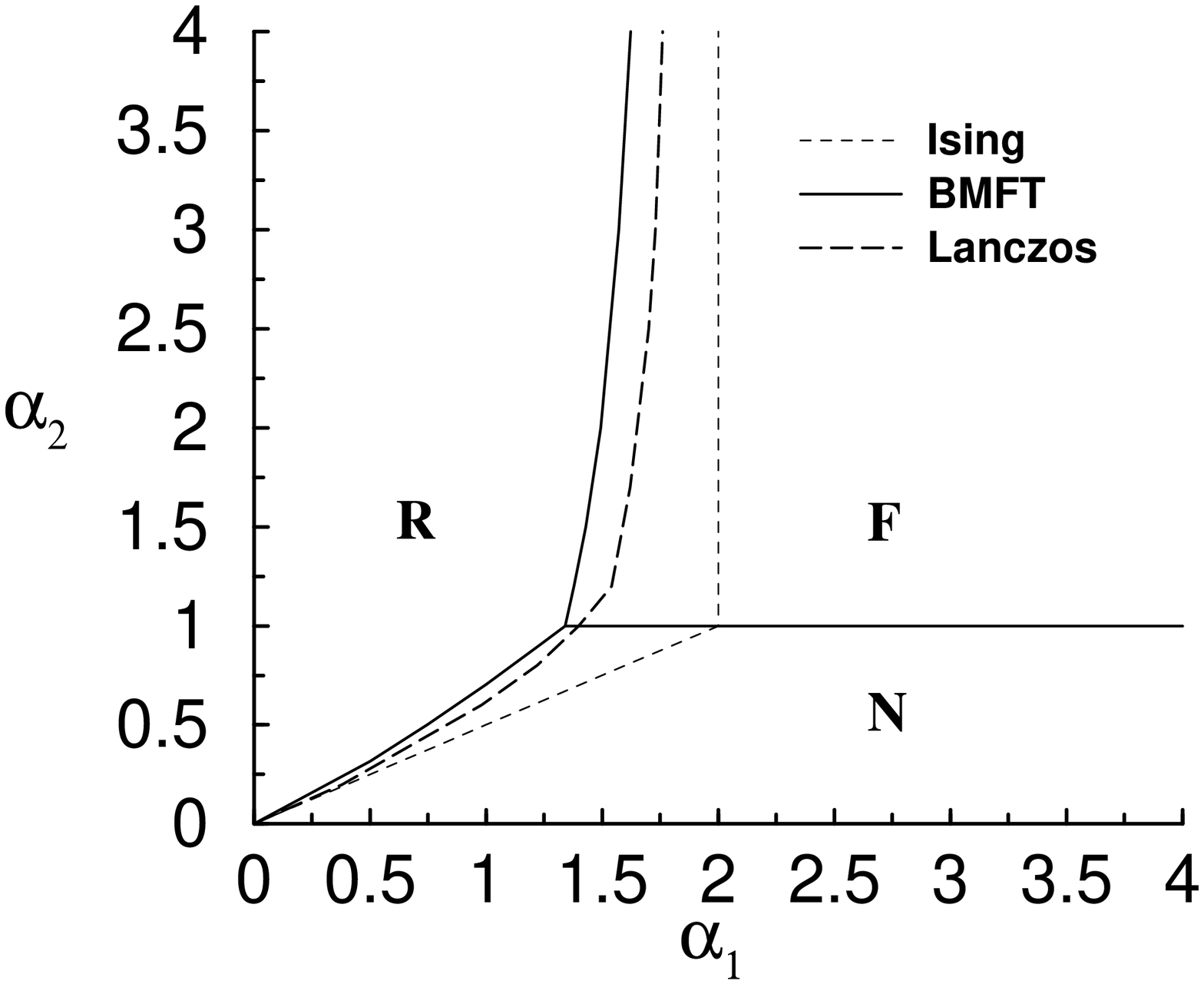}}(b)
\subfigure{
\includegraphics[scale=0.25]{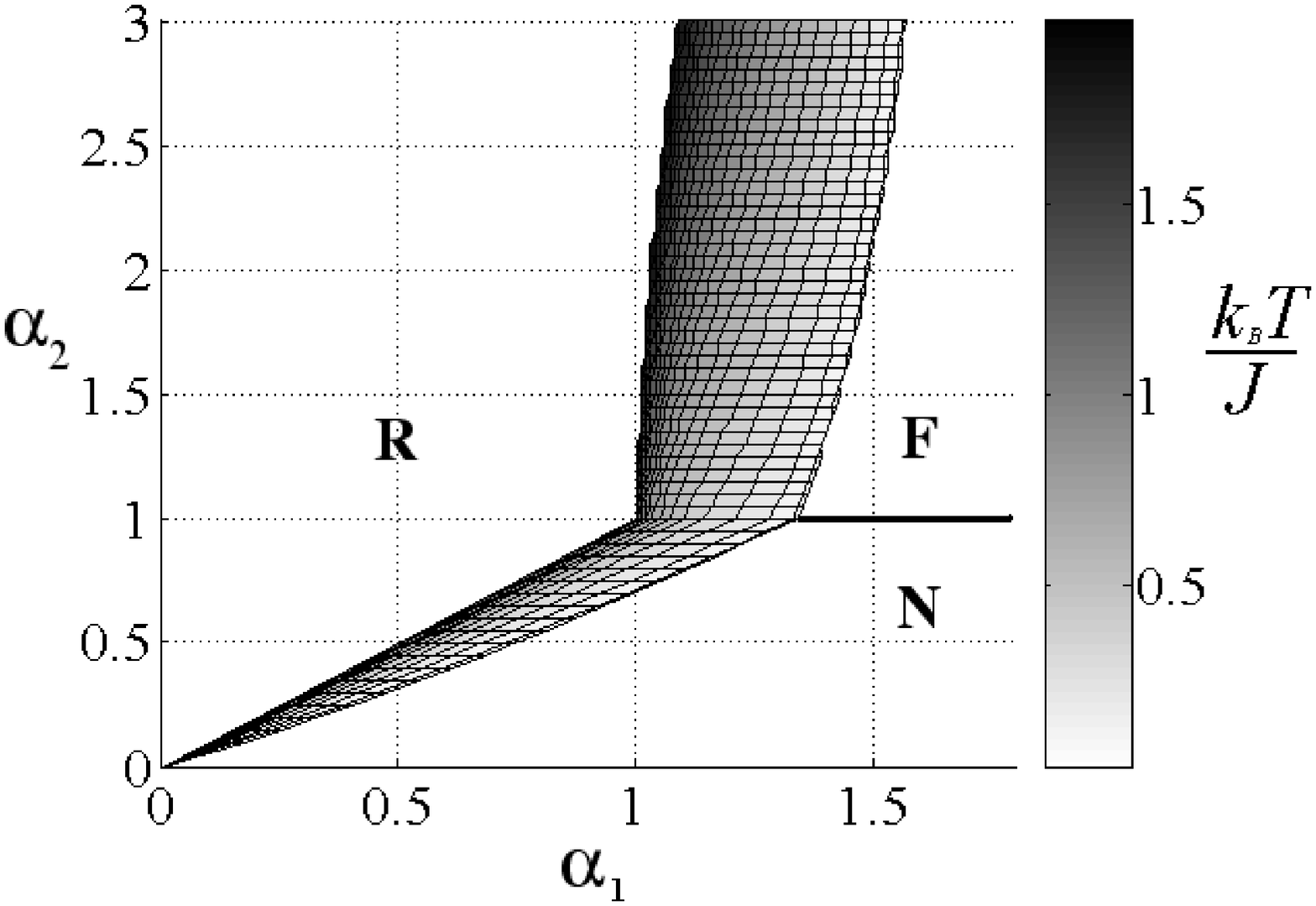}}		
\centering
\caption{The zero-$T$ (a) and finite-$T$ (b)
$(\alpha_1,\alpha_2)$-phase diagrams 
are shown. The zero-$T$ one is compared with the 
Lanczos-method~\cite{Jd6} and the Ising limit. The boundaries are between the N-type state (N), 
R-type state (R), and the F-type state (F). }
\label{fig:phase1}
\end{figure}

 The free (ground-state) energies of all three states 
are calculated as functions of the coupling constants and 
compared. From thermodynamic considerations the state 
with the lowest free energy is the stable one, and 
whenever free energies cross a phase transition takes 
place.  Since only the ratios of the couplings are 
important we define $\alpha_1 =J_{\perp}/J$ and 
$\alpha_2 = J_{\times}/J$. In this way we have 
obtained the zero and finite temperature phase 
diagrams which can be seen in Fig.~\ref{fig:phase1}. 
The agreement between the Lanczos method data and 
our results is very good, a fact that indicates that 
the present mean-field treatment is acceptable. The 
line at $\alpha_2=1$ is exact and its placement is 
a consequence of the Hamiltonian symmetry with 
respect to exchanging $J$ and $J_{\times}$. 
The quantum phase transitions (zero temperature) 
found here using BMFT are first-order ones for all 
values of $\alpha_2$. 
%Experimentally for a real 
%material, one can vary the pressure and hope that 
%the diagonal (or any other) coupling changes enough 
%so that the critical region is reached. 
As temperature increases the R-type state decreases 
in size. The sizes of the N-type and F-type phases 
increase with temperature. For any set of coupling 
values in the shaded region of Fig.~\ref{fig:phase1}b 
a classical (thermally induced) first 
order transition from the R-type state to one of the other states
occurs \cite{ramakko2007}.

\begin{figure}(a)
\subfigure{
\includegraphics[scale=0.20]{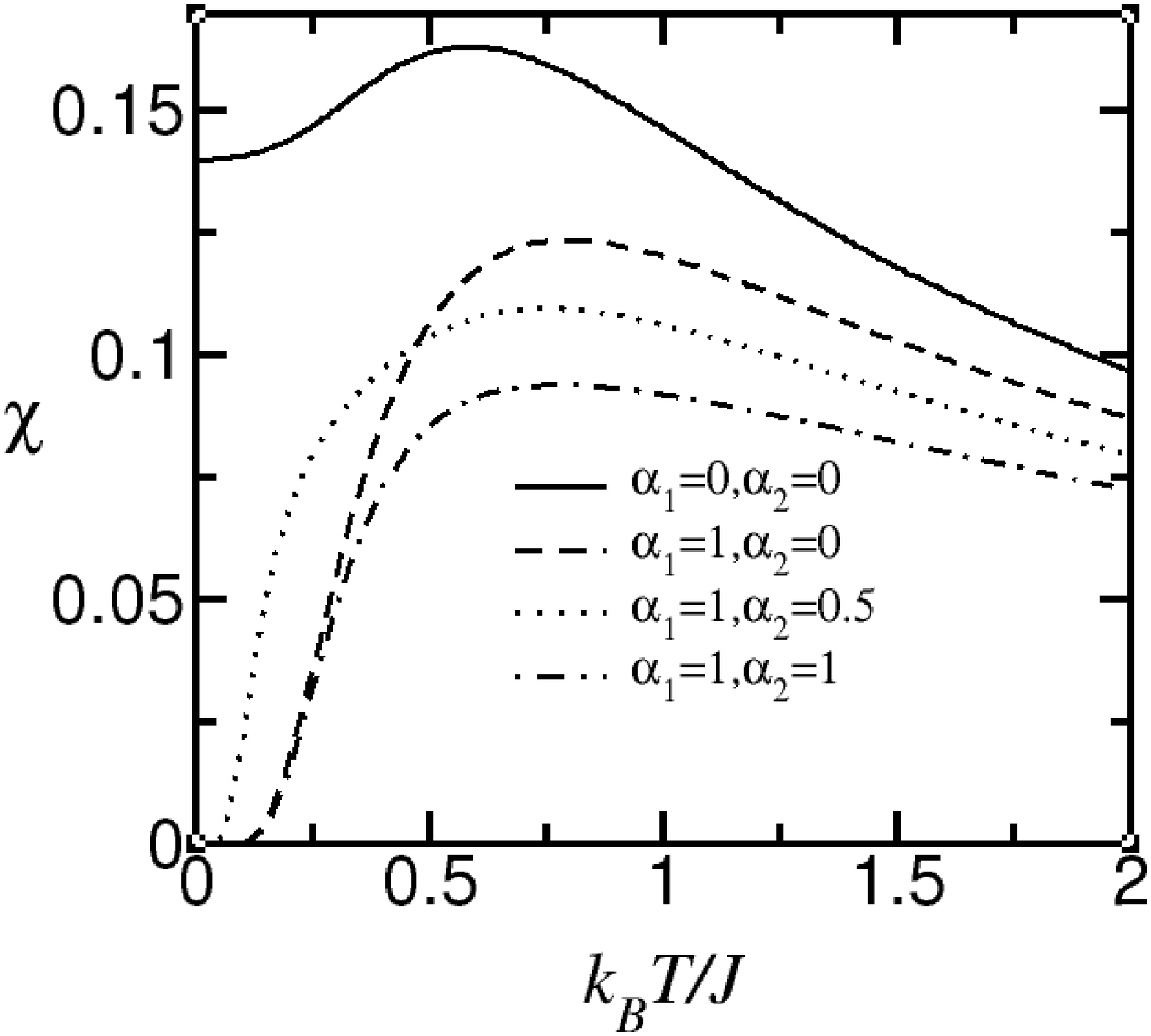}}(b)
\subfigure{
\includegraphics[scale=0.20]{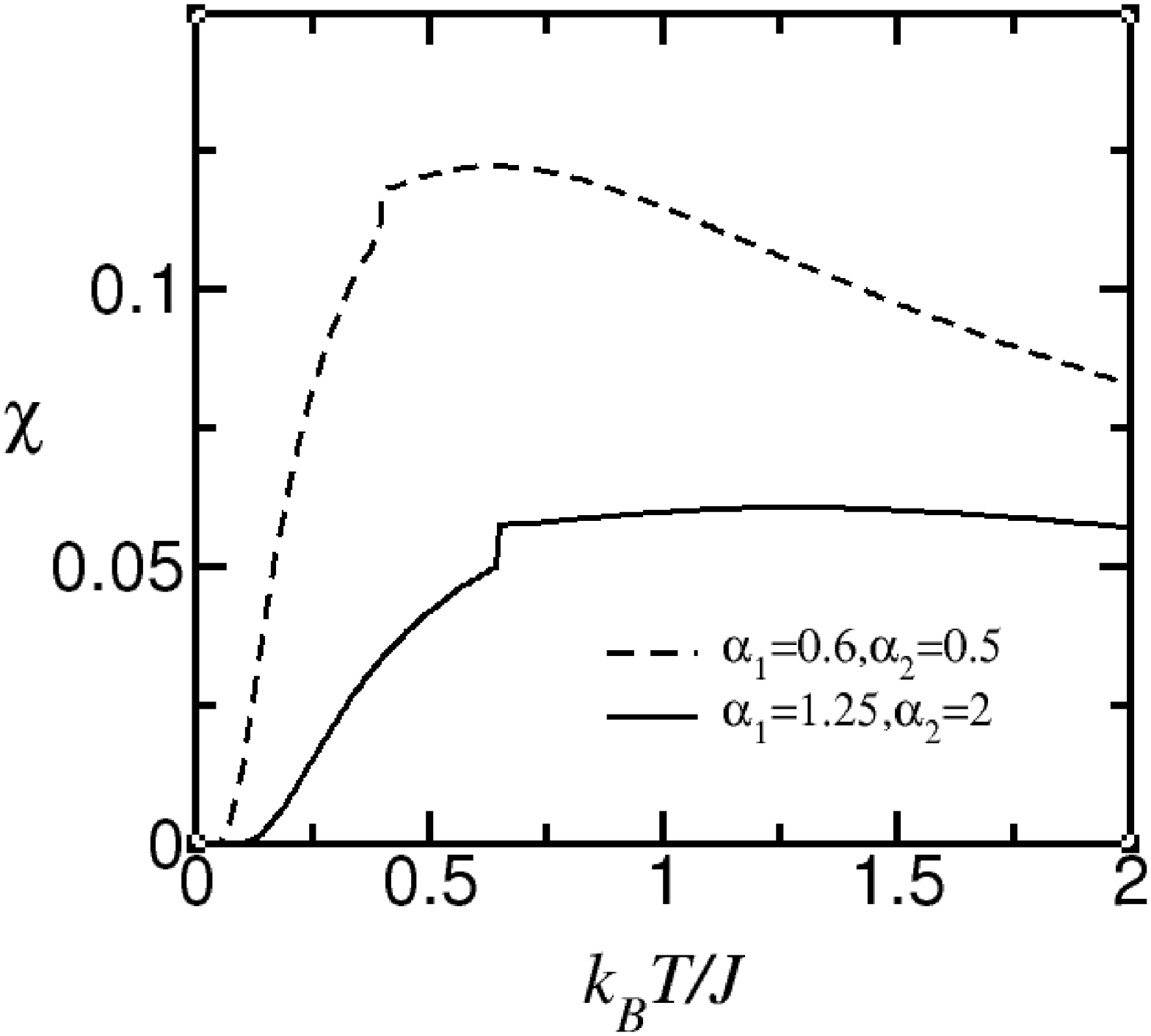}}		
\centering
\caption{\label{fig:suscept} The uniform spin susceptibility $\chi(T)$ 
is plotted as a function of temperature for $h=0$ for several coupling 
constants.}
\end{figure}

The uniform magnetic susceptibility $\chi(T)$ 
is calculated following the same method as Ref.~\cite{Azz5}. 
It is plotted as a
function of temperature in Fig.~\ref{fig:suscept} . 
The Heisenberg model on a chain
($\alpha_1=0,\alpha_2=0$) is gapless whereas when $\alpha_1\neq0$ 
and/or $\alpha_2\neq0$ an energy gap opens up in the 
low-energy excitation. This is why at zero 
temperature $\chi \neq 0$ only for the chain, and an exponentially
decreasing susceptibility is obtained for nonzero 
$\alpha_1$ or $\alpha_2$.
For $\alpha_1=0.6$ 
and $\alpha_2=0.5$ there is a sudden transition from the R-type 
state to the N-type state at $k_BT/J=0.39$. 
For example, for $\alpha_1=1.25$ 
and $\alpha_2=2$ there is a sudden transition from the R-type 
state to the N-type state at $k_BT/J=0.64$. 
This sudden transition results in a 
discontinuity in $\chi$. These findings remain to be confirmed
by others means.

\section{Conclusion}
\label{sec:Conclusion}

In this talk, quantum and classical critical behaviours in the 
frustrated antiferromagnetic two-leg ladder were presented.
The method of calculation, which is based on
the Jordan-Wigner transformation and the bond-mean-field 
theory was explained. 
The zero-temperature phase diagram of this system was explained. 
It exhibits three quantum phases, characterized all by an 
energy gap and absence of magnetic order. These states are labeled 
N\'eel-type, Rung-type and 
Ferromagnetic-type chain states. Our zero-$T$ results agree well 
with existing numerical data. When temperature increases 
for some sets of coupling values, the system 
undergoes a phase transition from the R-type state to the 
N or F-type state at a finite temperature. The finite temperature 
phase diagram was explained as well. In it, the size of the R-type 
state becomes smaller while the F-type state and the N-type state 
increase in size with increasing temperature. Our theory 
predicts a discontinuity in the spin susceptibility at a
finite temperature for the sets of couplings where the finite-$T$
transition occurs.

\section*{Acknowledgements}
We wish to acknowledge the financial support of the Natural 
Science and Engineering Research Council of Canada
(NSERC), and the Laurentian University Research Fund (LURF).

\end{document}